\documentclass{elsart}

\usepackage{epsfig}

\usepackage{amssymb}
\usepackage{bm}

\begin{document}

\begin{frontmatter}



\title{Thermal instability of an optically thin dusty plasma}


\author{Madhurjya P. Bora}\footnote[1]{E-mail~:~\tt mpb@guphys.cjb.net}

\address{Physics Department, Gauhati University, Guwahati 781~014, India.}

\begin{abstract}
We investigate the role of thermal instability, arising from radiative cooling
of an optically thin, dusty plasma, by linear stability analysis. The
corresponding isobaric stability condition for condensation mode is found to be
modified significantly by the degree of ionization and concentration of finite
sized, relatively heavy, and negatively charged dust particles. It has been
shown that though the fundamental wave mode is similar in nature to the one in
absence of dust particles, a new dust-wave mode can propagate in such a plasma.
It is conjectured that the presence of negatively charged dust particles may
considerably affect the stability of various astrophysical structures against
thermal instability, which can not be explained with the help of gravitational
instability.
\end{abstract}

\begin{keyword}
thermal instability \sep dusty plasma
\PACS 53.20.Ex \sep 52.30.-g \sep 98.58.-w
\end{keyword}
\end{frontmatter}

The formation and existence of a large number of astrophysical structures such
as interstellar clouds, solar prominence, localized structures in planetary
nebulae etc.\ can not be explained by means of conventional gravitational
instability \cite{parker}. However, it has been argued that thermal
instability may be a reasonably good candidate, with can accelerate
condensation, giving rise to localized structures which grow in
density by loosing heat, mainly through radiation \cite{parker,field}.
The first comprehensive analysis of thermal instability in a diffuse
interstellar gas is first given by Field \cite{field}. Subsequently, many
authors considered the process of thermal instability in different
circumstances e.g.\ in hydrogen plasmas \cite{defouw}, to explain formation of
clumpy gas clouds in two phase medium \cite{burkert}, and thermal instabilities
in photoioized interstellar gas \cite{corbelli}. Recently, Birk \cite{birk} has
considered thermal condensations in a weakly ionized hydrogen plasma.

In recent years, however, it has been realized that heavier dust particles
constitutes an ubiquitous and important component of many astrophysical
plasmas. \cite{birk,shukla,copp} including interstellar clouds, stellar and
planetary atmosphere, planetary nebulae etc. It is, thus of particular interest
to see the effect of dust particles on the role of thermal instability in
astrophysical situations. One of the obvious facts of introducing dust as a
component of plasma is the propagation of very low frequency compressional
modes \cite{rao}. We, in this paper, investigate thermal instabilities in a
multicomponent plasma with electrons, ions, and dust
particles, assuming that $a\ll d\ll\lambda_D$, where $a$ is the average dust
size, $d$ is the average inter-dust distance, and $\lambda_D$ is the Debye
length. We consider the dust particles to be negatively charged and cold,
whereas the ions can loose energy through radiation by de-excitation. We
further assume that electrons are isothermal. We introduce a heat-loss function
\cite{field} exclusive of thermal conduction, which takes care of the
radiative loss. For simplicity, we do not take into account the dust-charge
fluctuation, which usually has a stabilizing effect \cite{singh}. The resultant
equations are two-fluid magnetohydrodynamics (MHD) equations with Boltzmann
distribution for the electrons. The dust acoustic (DA) and dust-ion acoustic
(DIA) modes \cite{rao} are ruled out by virtue of `cold dust' assumption. We
show that the isobaric condition for thermal condensation mode is significantly
modified by the inclusion of dust dynamics. We also find the existence of a new
wave mode, solely because of the presence of the dust particles.

We begin by writing out the equations for a unbounded, weakly collisional
dusty plasma with a heat-loss function $\mathcal{L}(\rho,T)$, which represents
the rate of net loss of energy per unit mass, through radiation. We consider
the dust particles to be negatively charged and cold and the electrons obey
Boltzmann relation. We assume here that that ion-neutral collisional cross-section
is of the same order as that of the electron-neutral collisions, so that the 
electrons can be treated as Boltzmann distribution if $(T_i/T_e)
(v_{Te}/v_{Ti})\gg1$, and in practice this condition is satisfied in the cases
we are going to consider in this work. The equations are,
\begin{eqnarray}
\frac{\partial n_d}{\partial t}+\nabla\cdot(n_d)\bm{v}_d &=& 0,
\label{dust_cont}\\
m_d\frac{d\bm{v}_d}{dt} &=& -eZ_d\bm{E},
\label{dust_mom}\\
n_e &=& n_{e0}\exp\left(\frac{e\phi}{T_e}\right),
\label{boltz}\\
\frac{\partial n_i}{\partial t}+\nabla\cdot(n_i\bm{v}_i) &=&
-\nu_Rn_i+\nu_In_e,
\label{ion_cont}\\
m_in_i\frac{d\bm{v}_i}{dt} &=& eZ_in_i\bm{E}-\nabla p_i,
\label{ion_mom}\\
\frac{3}{2}n_i\frac{dT_i}{dt}+p_i(\nabla\cdot\bm{v}_i) &=&
\nabla\cdot(\chi_i\nabla T_i)-m_in_i\mathcal{L}(n_i,T_i),
\label{ion_temp}
\end{eqnarray}
where $\chi_i$ is the ion thermal conductivity, $\nu_{I,R}$ are ionization
and recombination frequencies, and the other symbols have their usual meanings.
The subscripts $e$, $i$, $d$ respectively refer to the electron, ion,
and dust populations. Equations (\ref{dust_cont}) through (\ref{ion_temp}) are
the continuity and momentum equations for dusts, Boltzmann relation for the
electrons, and continuity, momentum, and energy equations for the ions. The set
of equations is closed by the quasineutrality condition,
\begin{equation}
Z_in_i = n_e+Z_dn_d\label{quasi},
\end{equation}
where $Z_{i,d}$ represents the ion and dust charge numbers, respectively. Note
that we have taken into account the possibility of multiply ionized ions. The
ion pressure is given by $p_i=n_iT_i$ ($T_j$ are expressed in energy units) and
we have assumed that $\Gamma$, the ratio of specific heats for the ions to be
5/3. The recombination loss term in the ion-continuity equation represents the
electron-ion recombination on the surface of the dust particles \cite{kaw} and
$\nu_R\propto\pi r_d^2n_dn_ic_s^2$, where $r_d$ is the dust radius and
$c_s=(T_e/m_i)^{1/2}$ is the sound speed.

We consider now a small perturbation of the form $\exp(-i\omega
t+i\bm{k}\cdot\bm{r})$. The linearized expression for electron, dust, and ion
densities are given by
\begin{eqnarray}
\tilde n_e &=& \tilde\phi,
\label{pert_ne}\\
\tilde n_d &=& -\frac{\omega_d^2}{\omega^2}\tilde\phi,
\label{pert_nd}\\
\tilde n_i &=&
\frac{1}{Z_i}\left(\delta_e-\frac{\omega_d^2}{\omega^2}\delta_dZ_d\right)\tilde
\phi,
\label{pert_ni}
\end{eqnarray}
where we have used the linearized form of the Boltzmann relation for the
electrons Eq.(\ref{boltz}), dust-continuity equation Eq.(\ref{dust_cont}),
and the quasineutrality condition Eq.(\ref{quasi}). We have used
$\omega_d=kc_d$ and $c_d^2=Z_dT_{e0}/n_{d0}$. The quantities with subscript `0'
and `1' are equilibrium and perturbed quantities, respectively and
`$\,\tilde{}\,$' represents normalized perturbed quantities. We have normalized
the perturbed densities of the species by their respective equilibrium values
and perturbed electrostatic potential energy $e\phi_1$ by equilibrium electron
thermal energy, $\tilde\phi=e\phi_1/T_{e0}$. The quantities
$\delta_j=n_{j0}/n_{i0}$ are the ratios of equilibrium electron and dust
densities to that of the ions for $j=e,d$.

The linearized ion-continuity and momentum equations are given by
\begin{eqnarray}
-i\omega\tilde n_i+i(\bm{k}\cdot\bm{v}_{i1}) &=&
-\nu_R(\tilde n_d+\tilde n_i)+\nu_I\tilde n_e\delta_e,
\label{pert_ion_cont}\\
\omega(\bm{k}\cdot\bm{v}_{i1}) &=& \omega_i^2(\tau\tilde\phi
+\tilde p_i),
\label{pert_ion_mom}
\end{eqnarray}
where $\omega_i=kc_i$ is the ion-sound frequency, $c_i=(T_{i0}/m_i)^{1/2}$ is
the ion-thermal speed, $\tau=Z_iT_{e0}/T_{i0}$, and $\tilde p_i=p_{i1}/p_{i0}$
is the normalized perturbed ion pressure. Finally we write down the linearized
energy equation for the ions as
\begin{eqnarray}
-\frac{3}{2}i\omega\tilde
T_i+i\tilde\phi\left[\frac{1}{Z_i}\left(\delta_e-\delta_dZ_d
\frac{\omega_d^2}{\omega^2}\right)(\omega+i\nu_R)-i\left(\nu_I\delta_e
+\nu_R\frac{\omega_d^2}{\omega^2}\right)\right] &=& 0
\label{pert_ion_temp}\\
+\,\omega_\rho
\frac{1}{Z_i}\left(\delta_e-\delta_dZ_d\frac{\omega_d^2}{\omega^2}\right)
\tilde\phi+\omega_T\tilde T_i,
\nonumber
\end{eqnarray}
where we have defined $\tilde T_i=T_{i1}/T_{e0}$, $\omega_\rho=k_\rho c_i$,
and $\omega_T=(k_T+k^2/k_K)c_i$. Following Field \cite{field}, we
define the wave numbers $k_{\rho,T,K}$ as
\begin{equation}
k_\rho=\mathcal{L}_\rho\frac{m_in_{i0}}{c_i^3},\qquad
k_T=\mathcal{L}_T\frac{T_{i0}}{c_i^3},\qquad\mbox{and }
k_K=\frac{c_in_{i0}}{\chi_i}.
\label{k_defn}
\end{equation}
The first two are the wave numbers of isothermal and isochoric perturbations,
respectively and the third one is the reciprocal of mean free path of the
heat-conducting particles. It has been assumed throughout the
derivation that the perturbed ion thermal
conductivity $\tilde\chi_i=0$ and the equilibrium heat-loss
function $\mathcal{L}(\rho_0=m_in_{i0},T_{i0})=0$, where
$\mathcal{L}_{\rho,T}$ represents $(\partial\mathcal{L}/\partial T)_\rho$ and
$(\partial\mathcal{L}/\partial\rho)_T$, evaluated for the equilibrium state.

Using Eqs.(\ref{pert_ne})-(\ref{pert_ion_temp}), the dispersion relation for
this radiating dusty plasma can now be written as
\begin{eqnarray}
\left(1+i\frac{2}{3}\frac{\omega_T}{\omega}\right)\!\!\left[\frac{\omega^2}
{\omega_i^2}\frac{1}{Z_i}\left(\delta_e-\delta_dZ_d
\frac{\omega_d^2}{\omega^2}\right)\left(1+i\frac{\nu_R}{\omega}\right)
\right. &=& 0
\label{dr}\\
\left.-\,\tau-\frac{1}{Z_i}\left(\delta_e-\delta_dZ_d
\frac{\omega_d^2}{\omega^2}\right)-i\frac{\omega}{\omega_i^2}
\left(\nu_I\delta_e+\nu_R\frac{\omega_d^2}{\omega^2}\right)\right]
\nonumber\\
+\,\frac{2}{3}\left[\frac{1}{Z_i}\!\left(\delta_e-\delta_dZ_d
\frac{\omega_d^2}{\omega^2}\right)\!\left\{\frac{i}{\omega}
(\omega_\rho-\nu_R)-1\right\}+\frac{i}{\omega}
\left(\nu_I\delta_e+\nu_R\frac{\omega_d^2}{\omega^2}\right)\right],
\nonumber
\end{eqnarray}
where we have used the relation $\tilde p_i=\tilde n_i+\tilde T_i$.

Before numerically solving the dispersion relation Eq.(\ref{dr}), however, it
is instructive to analyze Eq.(\ref{dr}) under certain limits. In the limit of
collisionless electron-ion plasma with no dust particles and thermal radiation
and zero heat-conductivity, Eq.(\ref{dr}) reduces to
\begin{equation}
\omega^2=\omega^2_i\left(\frac{5}{3}+Z_i\frac{T_{e0}}{T_{i0}}\right),
\label{limit1}
\end{equation}
which is just the ion-sound wave modified by the ion-charge number. We now look
at the expansion of Eq.(\ref{dr}) \cite{field}, in the limit of weak collisions
($\nu_{I,R}\ll\omega$) and vanishing thermal perturbation i.e.\
$\omega_{\rho,T}\rightarrow0$, which, for the
condensation mode, is given by
\begin{eqnarray}
n_{\rm cond} &\simeq& \frac{2\delta_e}{(5\delta_e+3Z_i\tau)}\left[
\omega_\rho-\left(1+\frac{Z_i}{\delta_e}\tau\right)\omega_T
-(\nu_R-Z_i\nu_I)\right]
\label{limit2}\\
&&
+\,\frac{8\delta_e^4}{\omega_i^2(5\delta_e+3Z_i\tau)^4)}(3\omega_\rho+
2\omega_T)\left[\omega_\rho-\omega_T\left(1+\frac{Z_i}{\delta_e}
\tau\right)\right]^2
\nonumber\\
&& +\,\mathcal{O}\bigl(\omega_{\rho,T}^5\bigr),
\nonumber
\end{eqnarray}
where we have substituted $\omega=in$ and neglected $\omega_d$. The
characteristic velocity of the modes, we are interested, is $c_i$ and in a
plasma with cold dusts ($T_d\ll T_{i,e}$), as this one, $c_i/c_d\gg1$. For
example, in interstellar clouds, $n_i\sim10^{-3}$, $n_d\sim10^{-7}$, $T_e\approx
T_i\sim10\,\rm ^\circ K$ and with $Z_d\sim10^2$ and
$m_d\sim10^8m_i$, $c_i/c_d\sim10^3$, so that we can safely assume
$\omega_d\ll\omega$. To the leading order, we have the radiation condensation
mode if,
\begin{equation}
\omega_\rho+Z_i\nu_I>\nu_R+\left(1+\frac{Z_i}{\delta_e}\tau\right)\omega_T,
\label{cond1} \end{equation}
which reduces, in the collisionless limit to %
\begin{equation}
\omega_\rho>\left(1+\frac{Z_i}{\delta_e}\tau\right)\omega_T.
\label{cond2}
\end{equation}
Note that for zero thermal-conductivity, $\omega_T=k_Tc_i$ and the above
conditions are independent of $k$. The conditions given by inequalities
(\ref{cond1}) and (\ref{cond2}) are similar to the isobaric instability
condition given by Field \cite{field}, modified by collisions and dust
particles. We observe that the presence of a significant dust population
(smaller $\delta_e$) considerably modifies the instability criteria for the
condensation mode and has a stabilizing effect. It is interesting to note that
the degree of ionization has also a similar effect on the stability of the
condensation mode which becomes important in certain situations. For example,
in condensations observed in planetary nebulae, the thermal instability
is strongly associated with collisions of electrons with $\rm N^+$ and $\rm
N^{++}$ ions \cite{field}. It should, however, be noted that
the conditions (\ref{cond1}) and (\ref{cond2}) do not reduce to one given by
Field even in absence of dust particles, which is because our consideration of
two-fluid approximation.

\begin{figure}
\begin{center}
\epsfig{file=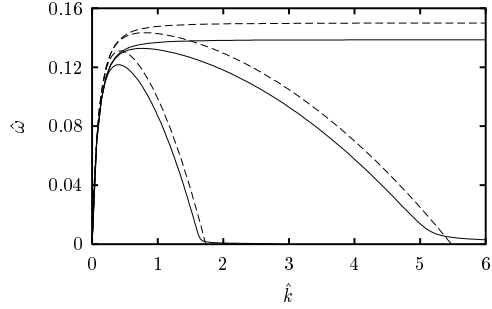,width=2.5in}
\end{center}
\caption{Normalized growth rate of the collisionless condensation mode
vs.\ wave number. The solid curves
are with dust particles whereas the dashed ones are without any dust. The
topmost curves are for $\hat k_K^{-1}=k_\rho/k_K=0$ and the others for
$\hat k_K^{-1}=0.01$ and 0.1, respectively.}
\end{figure}

The inequality (\ref{cond1}) also implies that the growth rate of the
condensation mode is independent of the wave number in the short wavelength
limit, when the thermal-conductivity is zero in opposite to a thermally
conducting plasma (finite $k_K$), where there is no condensation for
sufficiently larger $k$. Plots of the normalized growth rate
($\hat\omega=\omega/\omega_\rho$) versus normalized wave number ($\hat
k=k/k_\rho$) for the condensation mode are shown in Fig.1, in absence
of collisions. The growth rate of the mode is definitely smaller in
presence of dust particles, as we have guessed, but note the lengthening of the
cut-off wave number. In fact, the {\it knees} of the lower curves with finite
$k_K$ prove to a bifurcation point in a collisional plasma, as we shall see
shortly. The parameters for Fig.1 are $\hat k_T=k_T/k_\rho=0.2$, $Z_i=1$,
$Z_d=100$, $T_{e0}=T_{i0}=10\,\rm ^\circ K$, $m_d/m_i=10^8$,
$n_{i0}=10^{-3}\,\rm cm^{-3}$, and $n_{d0}=10^{-6}\,\rm cm^{-3}$. These values
are typical for interstellar clouds \cite{mendis}.

An approximation similar to Eq.(\ref{limit2}) for collisionless
wave modes is given by
\begin{eqnarray}
n_{\rm
wave} &=& -\frac{\delta_e}{(5\delta_e+3Z_i\tau)}\left(\omega_\rho+
\frac{2}{3}
\omega_T\right)\pm
i\omega_i\left(\frac{5\delta_e+3Z_i\tau}{3\delta_e}\right)^{1/2}
\label{limit3}\\
&& +\,\mathcal{O}(\omega_{\rho,T}^2).
\nonumber
\end{eqnarray}
The instability criteria for a growing wave mode is thus
$(\omega_\rho+2/3\omega_T)<0$, which is just the isentropic instability
criteria, given by Field \cite{field}. We can thus argue that the
principal wave mode is unaffected by the presence of dust
particles, but for a decrease in magnitude, dictated by the term $(5
+3Z_i\tau/\delta_e)^{-1}$, which proportional to $\delta_e/Z_i^2$. However,
there is one more wave mode which is not apparent from the above analysis. We
can detect the latter only by appropriately solving Eq.(\ref{dr}), numerically.
\begin{figure}
\begin{center}
\epsfig{file=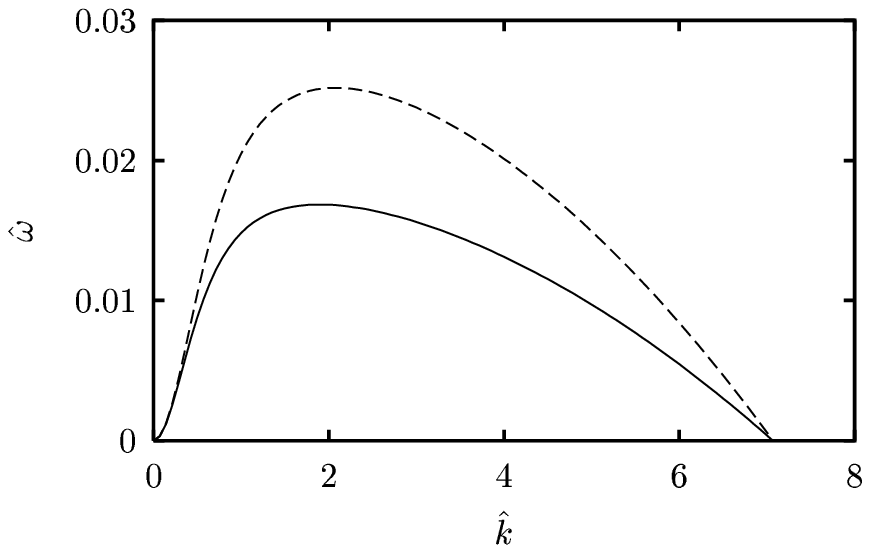,width=2.5in}%
\hfill%
\epsfig{file=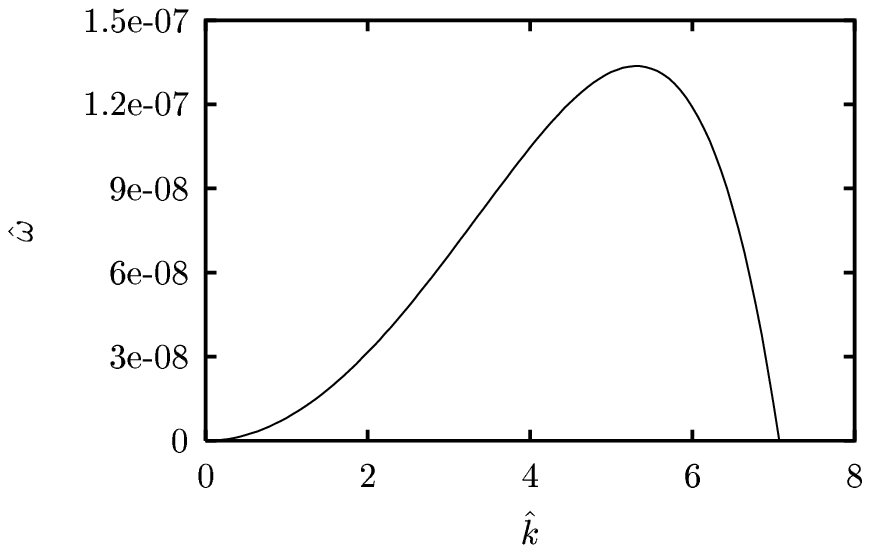,width=2.5in}
\end{center}
\caption{Normalized growth rates vs.\ wave number for the condensation modes
in absence of collisions. The first figure shows the usual wave mode present
irrespective of the dust particles. However, the growth rate in presence of
dust particles is considerably smaller (solid curve). The second figure shows
the dust-wave mode which is absent in a pure electron-ion plasma.}
\end{figure}
In Fig.2, we show the the growing-wave modes in a collisionless plasma with
considerable dust presence as in case of a planetary nebula \cite{mamun}.
There is a purely dust-wave mode (second figure in Fig.2) which is otherwise
absent in a pure electron-ion plasma. The growth rate of the mode is, however,
much smaller. Notice the cut-offs in all these curves. These figures are
with $\hat k_T=-2$, $\hat k_K^{-1}=0.01$, $Z_i=2$,
$Z_d=100$ and
$n_{d0}=10^{-5}\,\rm cm^{-3}$, all other parameters being same as in Fig.1. The
value of $\hat k_K$ is typical in planetary nebulae. We have chosen a large
negative value for $\hat k_T$ so as to drive the wave mode unstable (see
Eq.(\ref{limit3})). Note that a condensation mode will always accompany these
wave modes (not shown) as the condition for condensation instability is readily
satisfied in this case ($\hat k_T<0$).

\begin{figure}
\begin{center}
\epsfig{file=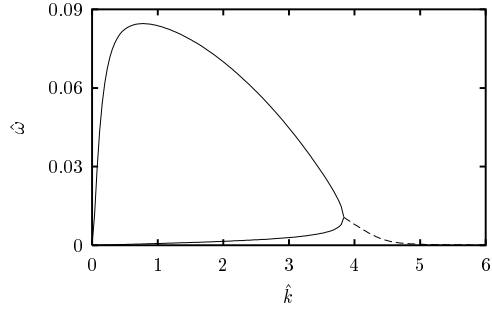,width=2.5in}
\end{center}
\caption{Growth rates in presence of collisions with $\hat k_T=0.2$,
$\hat k_K^{-1}=0.01$, $\nu_R/\omega_\rho=0.2$, $\nu_I/\omega_\rho=0$, and
$n_{d0}=10^{-6}\, \rm cm^{-3}$. Other parameters are same as in Figs.1 and 2.}
\end{figure}

In Fig.3, we show the growth rates of modes with collisional effects. The solid
curves represent condensation modes and the dashed one represents the wave
mode. We see a merger of the condensation modes to a growing wave mode at a
bifurcation point near $\hat k\simeq4.0$.

We should emphasize that we do not have an explicit expression for the
interstellar heat-loss function $\mathcal{L}(\rho,T)$ in general, because of
an wide variety of compositions of material in different interstellar regions.
Besides, several heating and cooling mechanisms may be operative simultaneously
and their dependence on local density and temperature may not be known exactly.
Delgarno and McCray \cite{delgarno}, in their classic review, have mentioned
about different cooling  processes in HI regions. A relatively concise
expression for the heat-loss function in photo-ionized interstellar gases with
high metallicity is given by Corbelli and Ferrara \cite{corbelli}. In the
simplest case, for radiative cooling, the heat-loss function can be assumed to
be proportional to $\rho^np^m$, where $n=5/2,
m=-1/2$ for free-bound transitions and $n=3/2, m=+1/2$ for free-free
transitions \cite{defouw,delgarno,talwar}. However, the exact form the
heat-loss function is not important at this point. As a prototype example, we
assume the heat-loss function used by Field \cite{field} for thermal
condensations in planetary nebulae, which is represented by
\begin{equation}
\frac{T_0\mathcal{L}_T}{\rho_0\mathcal{L}_\rho} =
\frac{\threehalf\hat k_K^{-1}T-(\half\bar T+\hat k_K^{-1})+\bar T/T}
{\bar T-\hat k_K^{-1}T}
\label{heat-loss}
\end{equation}
where $\bar T$ is the mean excess ionization temperature of $\rm N^+$ ions.
The temperatures are given in the units of $T_{\rm N^+}=21800\,^\circ\rm K$,
the mean excitation temperature of the $\rm N^+$ level. Applying the
instability condition (\ref{cond2}) in the collisionless case, we have, thermal
condensation if,
\begin{equation}
T_-(\bar T,\zeta)<T<T_+(\bar T,\zeta),
\label{cond3}
\end{equation}
where $\zeta=(1+1/\delta_e)$ and
\begin{eqnarray}
T_{\pm}(\bar T,\zeta) &=& \bigl[2\zeta+\hat k_K\bar T(2+\zeta)\pm
\bigl\{[2\zeta+\hat k_K\bar T(2+\zeta)]^2\bigr.\bigr.
\label{t_limit}\\
&& \bigl.\bigl.-\,8\hat k_K\bar
T\zeta(2+3\zeta)\bigr\}^{1/2}\bigr](4+6\zeta)^{-1}.
\nonumber
\end{eqnarray}
For simplicity, we have assumed $T_{e0}=T_{i0}$ and $Z_i=1$. The above
expression of $T_{\pm}$ reduces to usual one when $\zeta=1$ (single-fluid
approximation). For $\hat k_K=3/2$, $Z_d=100$, $n_{i0}=10^{-3}\,\rm cm^{-3}$,
$n_{d0}=10^{-7}\,\rm cm^{-3}$, and $\bar T=4.5$, we have $T_\pm=1.3293-2.542$,
which in absence of dust becomes $0.7248-3.7254$ \cite{field}. So, the presence
of dust provides a stabilizing effect against radiative condensations. However,
the range of instability shrinks to zero at $\bar T=3.902$, much higher than
1.974, which is for the dust-less case. This implies that for a planetary
nebula with a considerable amount of dust particles, to be stable against the
radiative condensation mode, it has to be at a much higher temperature. In
Fig.4, we show the range of unstable temperature with $\delta_e$. Note that a
lower value of $\delta_e$ corresponds to higher dust concentration. %
\begin{figure} \begin{center}
\epsfig{file=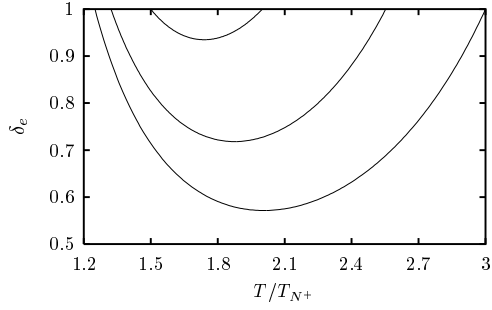,width=2.5in}
\end{center}
\caption{Range of instability in terms of temperature with $\delta_e$. The
value of $\bar T=4.0,4.5$, and 5.0, respectively for the upper, middle, and
lower curves.}
\end{figure}

To conclude, we have found that the instability criterion for the thermal
condensation mode is significantly affected by the presence of dust
particles, which can severely alter the stability regimes of many astrophysical
condensations where a significant number of dust particles are present. The
fundamental wave mode is similar in both cases with and without
dust particles, though we now have another wave mode, which is due to
the presence of dust particles. Finally, we have shown that the conventional
stability regimes of many astrophysical condensations may be considerably
changed.

\section*{Acknowledgement}
I would like to acknowledge the hospitality during my stay at The Abdus Salam
ICTP, Trieste, Italy, where a part of this work has been carried out.


\begin{thebibliography}{00}




\bibitem{parker} E.N.\ Parker, Ap.\ J.\ 117 (1953) 431.

\bibitem{field} G.B.\ Field, Ap.\ J.\ 142 (1965) 531.

\bibitem{defouw} R.J.\ Defouw, Ap.\ J.\ 161 (1970) 55.

\bibitem{burkert} A.\ Burkert, D.N.C.\ Lin, Ap.\ J.\ 537(1) (2000) 270.

\bibitem{corbelli} E.\ Corbelli, A.\ Ferrara, Ap.\ J.\ 447 (1995) 708.

\bibitem{singh} R.\ Singh. M.P.\ Bora, Phys.\ Plasmas 7(6) (2000) 2335.

\bibitem{birk} G.T.\ Birk, Phys. Plasmas 7(9) (2000) 3811.

\bibitem{shukla} P.K.\ Shukla, G.T.\ Birk, G.E.\ Morfill, Phys.\ Scr.\ 56
(1997) 299.

\bibitem{copp} A.\ Copp, G.T.\ Birk, P.K.\ Shukla, Phys.\ Plasmas 4 (1997)
4414.

\bibitem{rao} N.N.\ Rao, P.K.\ Shukla, M.Y.\ Yu, Planet.\ Space Sci.\ 38 (1990)
543.

\bibitem{kaw} P.K.\ Kaw, R.\ Singh, Phys.\ Rev.\ Lett.\ 79 (1997) 3.

\bibitem{mendis} D.A.\ Mendis, M.\ Rosenberg, Ann.\ Rev.\ Astron.\ Astrophys.\
32 (1994) 419.

\bibitem{mamun} A.A.\ Mamun, P.K.\ Shukla, Phys. Plasmas 7(9) (2000) 3762.

\bibitem{delgarno} A.\ Delgarno, R.A.\ McCray, Ann.\ Rev.\ Astron.\ Astrophys.\
10 (1972) 375.

\bibitem{talwar} S.P.\ Talwar, M.P.\ Bora, J.\ Plasma Phys.\ 54(2) (1995) 157.

\end{thebibliography}
\end{document}